# Engineering Autonomous Driving Software

Christian Berger and Bernhard Rumpe

**Abstract** A larger number of people with heterogeneous knowledge and skills running a project together needs an adaptable, target, and skill-specific engineering process. This especially holds for a project to develop a highly innovative, autonomously driving vehicle to participate in the 2007 DARPA Urban Challenge. In this contribution, we present essential elements of a software and systems engineering process to develop a so-called artificial intelligence capable of driving autonomously in complex urban situations. The process itself includes agile concepts, like a test first approach, continuous integration of all software modules, and a reliable release and configuration management assisted by software tools in integrated development environments. However, one of the most important elements for an efficient and stringent development is the ability to efficiently test the behavior of the developed system in a flexible and modular system simulation for urban situations both interactively and unattendedly. We call this the simulate first approach.

## 1 Introduction and Motivation

Innovative research is often centered around interesting challenges and awards. The airplane industry started off with awards for the first flight over the British Channel as well as the Atlantic Ocean. The Human Genome Project, the RoboCups, and the series of DARPA Grand Challenges for autonomous vehicles serve this very same purpose to foster research and development in a particular direction. The 2007 DARPA Urban Challenge took place to boost development of unmanned vehicles for

Christian Berger
RWTH Aachen University, Department of Software Engineering, Ahornstraße 55, 52074 Aachen,
e-mail: mail@christianberger.net

Bernhard Rumpe
RWTH Aachen University, Department of Software Engineering, Ahornstraße 55, 52074 Aachen,
e-mail: rumpe@se.rwth-aachen.de

urban areas. Although there is an obvious direct usage for DARPA, there will also be a large number of spin-offs in technologies, tools, and engineering techniques, both for autonomous vehicles, but also for intelligent driver assistance systems.

Such a system needs to be able to understand the situation around the car, evaluate potential risks, and help the driver to behave correctly, safely, and, in case it is desired, also efficiently. These topics do not only affect ordinary cars, but also buses, trucks, convoys, taxis, special-purpose vehicles in factories, airports, mines, etc. It will take a while before we will have a mass market for cars that actively and safely protect the passenger and their surroundings from accidents in all situations but their development is on its way.

Intelligent functions in cars are obviously complex systems. For a stringent deadline-oriented development of such a system it is necessary to rely on a clear, usable, and efficient development process that fits the project's needs. Furthermore, changing requirements and enhancements of technologies need to be incorporated into the development effectively. This kind of situation is well known in business and web-based software development. Therefore, that industry has developed appropriate methods and process frameworks to handle this kind of projects.

Among a number of agile development processes, Extreme Programming (XP), Scrum, and the Crystal family of processes are the most prominent. However, these development processes are dedicated to software only and seem to not support traditional engineering processes properly at first sight, which often include the development and setup of embedded systems. Therefore, a compromising adaptation is necessary that addresses the needs of both worlds. Moreover, a coherent, efficient, and automatable tool suite is inevitable for a goal-oriented development project. We describe an efficient tooling infrastructure necessary for such a development project in Sec. 4. This tooling relies on a process for informal requirements to direct coding approaches which we describe in Sec. 2.

For quality management, quite a number of actions are taken. Among others, the steady integration, the requirement for continuously running code, and the regular integration into the vehicle are part of the plan. Most important, however, and this over time became a more and more obvious advantage in the "CarOLO" project which itself is described in Sec. 3, are the automated and unattended tests that have been developed for all parts and integration stages. Demanded by the test first development approach, tests for functions and algorithms are coded in a parallel manner to the actual code. The technology and measurements to enable this testing process are described in greater detail in Sec. 4 as well.

The fulfillment of requirements for a safe autonomous car needs to be very intensively tested. This is to a large extent achieved by running the software in a *virtual test-bed* for avoiding race conditions on valuable resources like the real vehicle. The system simulation acts like real surroundings for the software by producing information such that it thinks it was on the street and decides accordingly. This real world system simulation was developed not only to test the requirements, but also to allow the system to "virtually drive" through hundreds of different situations, which may additionally contain other cars. The system simulation therefore allows us to test–and especially to automatically re-test–the behavior of the soft-

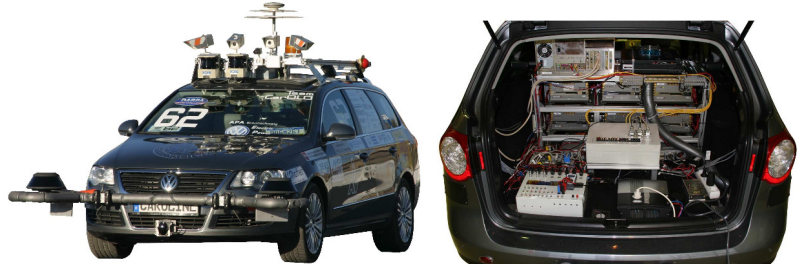

(a) For realizing an overlapping and thus a redundant field of view around the car to perceive the vehicle's surroundings, different sensors like single and multi-layer laser scanners, radars, and cameras were used.

(b) All data gathered from the sensors is transferred to Caroline's trunk, where different computers were used to process the data to derive driving decisions.

Fig. 1: A front and rear view of the vehicle "Caroline", an autonomously driving vehicle for urban environments (based on [13]).

ware in a regression testing manner without any possible harm. We can not only test for positive situations, but also for negative and physically impossible situations of the surroundings and obstacles. Furthermore, the system simulation can be used in an interactive mode to understand a vehicle's behavior. This is one of the basic elements for actually being able to develop our autonomously driving vehicle as shown in Fig. 1, called "Caroline" in the "CarOLO" project, in time to participate in the 2007 DARPA Urban Challenge. The technical aspects of the system simulation used in the CarOLO project are described in [2] which founded the basis for this article. However, in Sec. 6, an improved successor is outlined.

## 2 Overview of an Efficient Software Engineering Process

Whenever things become complex, they need to be decomposed into smaller pieces. Also when the development of a product is complex, not only the product, but also the development activity needs to be structured to become manageable. Both in the areas of software development and systems engineering, a variety of such processes exists. An appropriate process can be chosen based on the needs of the project in terms of complexity, criticality, and urgency of the product. However, due to the different nature of "virtual" software vs. physically existing hardware, these development processes differ greatly. Nowadays, this process gap is a constant source of problems. Before we look at the process used for the development of Caroline, we highlight a few of those distinctions.

Due to its immaterial nature, it is obvious that software can more easily be reorganized and evolved than physically existing hardware. Therefore, software devel-opment processes can be iterative and incremental to allow an evolutionary evolvement towards a stable, robust, and mature result. Recent processes like XP, Scrum, or RUP advocate iterations in all stages with various durations. Iterations are necessary to continuously evaluate the current state of the customer's needs on the one hand. On the other hand, iterations enforce to continuously integrate existing artifacts to a *functional* system. Thus, iterations are a vital part to deal with the necessary improvement of existing software, changing requirements, and an incremental process that decomposes software into smaller pieces. The more innovative software products are, the less predictable a software development process is. In such a context many small iterations are preferable over a few larger ones. More and smaller iterations allow the software management to adjust priorities and to lead the project more successfully.

We believe that in integrated software and systems development projects with a heavy software part an iterative software development process needs to be used instead of the traditional engineering process. The best approach is to decouple subprojects as much as possible such that the individual processes can be followed in subprojects. A set of high-level milestones connect the subprojects and ensure a coherent process on the top, but within each subproject different forms of processes are in use. Furthermore, it is inevitable to decouple iterative software development from hardware e.g. through virtualization. Looking into the software development process, we find that these many small iterations strongly imply a number of other development practices that are necessary to ensure progress and quality. However, most important is a continuous integration of the software pieces. Experience shows that software simply cannot be developed independently and integrated later, even if the interfaces are defined as precisely as possible, because patchwork inevitably must be done during integration that invalidates earlier development work. With modern tools for version control and configuration management, continuous integration can be achieved rather easily. The most tricky and hard to achieve part is to ensure that all team members are committed to a continuous integration process. This is because software must be developed in a collaborative fashion, where no code ownership exists and everybody releases and integrates his software at least several times a day.

Disciplined use of versioning is the basis for the next important process element, namely automated, integrated, and unattended testing. Testing is by far the most important technique for quality assurance and comes in many different flavors, beginning with unit tests to integration tests up to full system tests. Incremental software development requires periodic testing to test artifacts as well as their integration as already mentioned earlier. The "testing trap", which denotes unnoticed software bugs due to inconsequential software testing, can only be escaped through automated replaying of tests, also called regression testing, for each single increment. The important part of automated testing is not finding appropriate tests, but the techniques that run the current code to determine efficiently whether some property of the code is correct or wrong and without humans to run the test or interpret the results. As known from unit tests, automation helps each developer to know whether an error was introduced to the code, in which iteration it was introduced, and in which part of the software. Error detection thus becomes much easier and after detecting an error, an identification of the location is relatively easy within a small area.

A very important aspect of automated testing is that it cannot only be used to test each iteration, but even more every single version. This means each development subtask, may it be a 10-minute bugfix or a 3-hour development of an algorithm, can easily be checked against all available tests at no additional cost of manpower. We therefore integrated the automated testing infrastructure with the version control system: Each version which was committed to the versioning server triggered automatically a testing process, where all available tests are run and feedback in the form of the number of failed tests with detailed information is given. Our experience is that in the long run this kind of quality assurance helps very much to foster an efficient development of software. Initially however, it needs a lot of discipline. Furthermore, appropriate tooling infrastructure is inevitable to make the developers accept this discipline.

Even more discipline is necessary to start with developing tests first for software elements as there should not be any untested functions in the system at any time. Fortunately, developers can later enjoy seeing automatic tests run through to status "green" at every new version. Testing also needs to deal with configuration variations-a popular C/C++ problem allows code to behave differently on different platforms. The test first approach can help to identify problems early providing a so-called automated build farm for various platforms.

When developing autonomously driving cars, automated testing has a number of challenges to tackle. First of all software is integrated and closely linked with hardware, such as sensors and actuators, and through them to the surroundings. Appropriate abstractions for different parts of the software and the hardware are necessary to run tests efficiently. For the "intelligent part" of the software, it is not necessary to run tests based on full sensory input, but to provide distilled, aggregated information about possible obstacles as well as the pathway to drive through them. A highly important abstraction for efficient test automation is to replace the real hardware by a simulation. A simulation of the hardware allows automated tests on ordinary computers and is thus available for each developer independently. As all physical elements are simulated, it furthermore allows decoupling the time for running a software test from the real time. This allows to run a complete city traversal in a few seconds. We are thus able to run thousands of tests for every new version each night. As a prerequisite we needed to develop a testing infrastructure that:

- allows us to define various constellations and subsystem configurations of the software parts to be tested,
- provides a software test-bed to probe the software under test and to understand whether it behaved correctly,
- is capable of providing physics-based behavior of the controlled vehicle as well as the urban surroundings correctly and in sufficient detail, and
- allows us to easily define new tests including automated evaluation of the test results.

Of course, tests that deal only with the simulation software are not enough to ensure a robust automotive system. Therefore, a continuous deployment of the software subsystem into the car and a (re-) run of a sufficiently large set of tests in the vehicle are inevitable. For that purpose, we have an additional test team that runs the fully equipped car in various traffic situations. The test team gets software releases in regular iterations and checks the car's abilities against the requirements from traffic rules, DARPA Urban Challenge rules, and story cards mapping those requirements into concrete driving missions which is outlined in the following. However, a stringent software testing process considerably reduces the amount of time necessary to fully run the hardware-in-the-loop (HIL) tests. Furthermore, the complexity of traffic situations necessary to do the test usually requires several people in each test run. For example, any single test that deals with a correct behavior in a junction lasts for about at least 15 minutes to setup, run, and check the resulting behavior. It involves several other cars and people to produce appropriate junction situations. Multiplied by the variety of junctions and the many possibilities of cars coming in different directions, this would take far too long to actually run all of them in reality. So quite a number of junction situations are tested only in the system simulation. The possibility to rerun those situations efficiently and automatically is important to ensure the stringent and effective development process needed.

Another important issue to be taken care of from the beginning is to organize the software in appropriate subsystems and components, to define the technical interfaces, and to take additional actions so that the software can be developed and tested independently. Only if the software obeys a number of best practices and design patterns it is possible to test the software efficiently. For example we can decouple the time a test takes from the time the tested software thinks it runs in, if the software does not directly call the operating system about the current time or even worse, counts itself, but uses an adapter interface. Similar techniques are necessary, if outside software needs to be incorporated that does not have a testable software architecture, neighboring systems are not part of the testable subsystem, or sensors and actuators come into play. Architectural testing patterns helps to develop software that can be tested in various configurations and parts individually e.g. mock objects or abstract factories for object construction.

Moreover, we have adopted some more practices, e.g. from Extreme Programming, beyond short iterations. For example, one successful organizational and structuring tool were story cards. A story card describes briefly and explicitly the goals for a development iteration and thus leads to a useful, efficient, and focused structuring of the requirements and also the development project. Accompanied with a definition of measurable goals for every task, these story cards allow the developers to understand and measure progress of development.

Having highlighted a number of special issues that arise when integrating an agile software and a classic engineering process, we note that classic engineering and software engineering indeed have different development cultures. It takes a while until both cultures efficiently work together, but when properly integrated, the resulting output is tremendous and of very good quality.

## 3 The CarOLO Project

The CarOLO project was carried out from May 2006 until December 2007. In this project, the autonomously driving vehicle named "Caroline" was developed which participated in the 2007 DARPA Urban Challenge. The contribution was among only ten other vehicles from initially more than 85 vehicles from all over the world which achieved the Urban Challenge Final Event.

### *3.1 The 2007 DARPA Urban Challenge*

The 2007 DARPA Urban Challenge was the third major competition and the successor of the 2004 & 2005 DARPA Grand Challenges [6, 22]. Compared to those elder series in which the competitors had to develop a vehicle which was capable of driving autonomously through an a-priori unknown and rough terrain using a given digital map, the Urban Challenge lifted the requirements significantly.

In the Urban Challenge, not only a digital map had to be processed to reach a given set of destinations, but even more the vehicle had to deal with dynamic vehicles for the first time. These vehicles were controlled both by human safety drivers and by other "robot" competitors as well. Moreover, the car had to obey logical traffic rules which define priorities at intersections or set speed limits for example. Furthermore, the car had to park without colliding with other already parked vehicles in a parking lot.

The semifinal and the final itself took place at the former George Airforce Base in Victorville, CA. For the challenge, DARPA had rebuilt that area to simulate an urban environment with junctions, intersections, and parking lots.

For participating in the challenge, DARPA provided two possibilities: The first track called "Track A" was explicitly funded by DARPA, and the second track called "Track B" was not funded at all and for nearly everyone who could provide at least an American team leader for the project. The CarOLO project participated in the latter track.

Both tracks consisted of a multi stage participation process. The first major stage was to provide a video demonstrating the vehicle's capabilities at an early stage. In that video, the vehicle had to follow a road using the available lane markings and to stop right before an obstacle blocking its own lane. After a short while, the vehicle had to start a passing maneuver and merge into its own lane again to finish the course.

Using the video, DARPA selected teams for a closer inspection at the so-called "Site Visits" at performer's site. Due to the fact that the CarOLO project was initiated in Germany, the team and the vehicle had to be shipped to the United States for demonstrating its capabilities in June 2007. Therefore, our team cooperated with the Southwest Research Institute (SwRI) in San Antonio, TX.

The site visit demonstration started with an inspection of the emergency control system. Next, the vehicle had to follow its lane, pass a stationary vehicle as already shown in the video, show its U/K-turn capabilities, and behave correctly at intersections. Using this multi-stage participation process, DARPA could enforce a high quality for the semifinal.

For the semifinal which took place in Victorville, CA as mentioned before, only 35 teams were qualified after the site visit. The semifinal, which was called "National Qualification Event", provided three independent test areas which tested different aspects of the vehicles, respectively.

The first test area, "Test Area A", provided an eight shaped course to test correct merging at T-junctions. Therefore, the vehicle had to find a gap in the moving traffic to merge into. The second test area, "Test Area B", tested a vehicle's capabilities to pass safely any obstacles at the side of a road and to park the car in a parking lot. The last area, "Test Area C" re-tested correct behavior at intersections and dynamic re-planning in the case of blocked roads.

Based on the vehicles' performance in the semifinal, only eleven teams qualified for the 2007 DARPA Urban Challenge Final Event which took place on November, 3rd [16].

### 3.2 Team CarOLO's Caroline

The vehicle already shown in Fig. 1 was named "Caroline". The vehicle is a 2006 Volkswagen Passat Station wagon which provided the necessary technical possibilities like drive by wire to realize our software and system architecture. For a technical explanation with greater details we refer to [1] and [13].

In Fig. 2, the layered software architecture of Caroline is depicted. The architecture consists of several only loosely coupled modules for solving different tasks like detecting lanes, fusing incoming new sensor data, or updating the currently planned route. The entire system realizes the so-called *pipes-and-filters* pattern where one module processes incoming data for producing outgoing data which is the new incoming data for the following module. Thus, the task of driving autonomously is split into smaller pieces obeying the "separation-of-concerns" rule supported by a self-developed software framework.

The software framework provides rudimentary system services like concurrency and communication. Using this framework and the layered pipes-and-filters architecture, all modules could simply be reused for system simulation. Technical details for the system simulation used in the CarOLO project can be found in [2]. In Sec. 6, a new approach is presented for efficiently designing, using, and unattendedly executing system simulations which avoids some conceptual design problems from the software framework used in Caroline.

Fig. 2: System architecture for the autonomously driving vehicle "Caroline" (based on [13]).

## 4 Tools for an Efficient Software Engineering Process

Compared to any other development of complex software-intensive systems, the software development process described in Sec. 2 can only be accomplished if appropriate tooling infrastructure is available. This is important because to participate in the Urban Challenge, the complete software and hardware system has to be developed on a very tight schedule because there are no negotiations at all. For a successful participation, both efficiency and quality of actions have to be balanced wisely. These special circumstances led to the organizational implementation of the agile software and system engineering process based on Extreme Programming. Clearly, a modern IDE like Eclipse is used for direct code development in C++ and MAT-LAB/Simulink for the control algorithms. However, as MATLAB/Simulink does not scale for complex data structures or supports properly the design and implementation of software architectures, most of the code by far was written in C++. Therefore, in the following, we concentrate on the C++-part of the software development.

For planning purposes, plain mathematical functions are used to understand the algorithms, and UML class and deployment diagrams as well as statecharts are used for the software architecture as well as the important state-based behaviors and data structures. However, these diagrams are used for discussion and serve therefore as documentation only.

As described, milestones are centered on story cards as shown in Fig. 3 that serve as a definition of measurable goals. These measurable goals are the base for a consistent test process and its tooling infrastructure that is described in the following. The whole team is distributed in several locations and team members sometimes work in different time zones. For a uniform understanding of progress, a single source of information is necessary for collecting tasks, tracking bugs, and publishing progress. Using Trac as an integrated and easy to use web based portal for the complete software and system development process enables the team to track changes to the software over time, and evaluate the actual state of the software generated by the back-end tool chain. As mentioned above, every story card is virtually available for every team member at any time to see the most important aspects for the current and next development iteration. In addition to the description of the next iteration, a list of tasks, each with a metric of its completion and a list of open bugs are available for every virtual story card.

For achieving the goals described in a story card, among other things, system simulations as described in Sec. 6 can be used by developers to test their code. In an initial step, the requirements of the story card are translated into an initial number of tests which describe a *scenario* even before the software is developed. After fulfilling the requirements in the system simulation the software is put into operation on the real hardware.

To manage parallel development as well as different configurations in the system simulation and real hardware, a release and configuration management tool based on Subversion and FSVS [21] is used. This allows us to progressively enhance the software development in small increments, while at the same time the reload of older, stable versions for demonstrations and events, where a stable and demon-strable software version needs to be loaded on the car is supported. Furthermore, the configuration management has to ensure that hardware changes fit to the loaded software releases. As usual, all seven car computers are not only backed up in their own configuration, but also version-controlled.

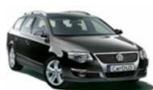
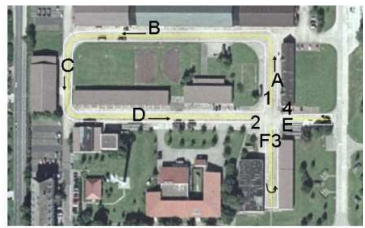

Fig. 3: Story card which describes the requirements in an intuitional manner for one scenario. This card is inspired by Volère's requirements' specification. On the first line, a unique identifier for this scenario is provided. On the left hand side, some information like a headline, date, and location are defined. These information are followed by a detailed description of the behavior which is expected from the vehicle to fulfill this scenario. On the right hand side, an image illustrating the scenario and its situations is depicted (for our example an aerial image of our testing site which is located on former barracks from the German Forces is shown). In the bottom area, the natural description is split into detailed and identifiable *functional* requirements for every module. Further columns are used for responsibilities and daily fulfillment rates; the actual milestone is indicated by the last green column.

Using FSVS as a version control system for filesystems enables the team to simply and safely test new software versions and to maintain the integration between parallel developments as well as tracking of open issues and potential bugs which were found during real vehicle tests. Based on version control the independent test team has the ability to retrieve specific software releases that the development team wants to be tested. This further decouples testing and development and allows more parallelization and thus increases efficiency. All car computers are consistently restored to the specific software release and a detailed test process based on the measurable goals of the virtual story cards can be carried out rather efficiently. In particular, bugs and behavioral issues can be recorded in such a way that they can be replayed and analyzed in detail if necessary. Both the development and the test teams can simply restore the development state of the car in the desired software release by switching every car computer to the appropriate revision using one simple command. Instead of FSVS, more recent filesystems like ZFS or Btrfs can be used for a similar purpose.

The combination of virtual story cards and a consistent release and configuration management enables the team to safely develop and test potentially dangerous new software functions without breaking an already running software system on the vehicle. Furthermore, the list of open or closed tasks allows the project management to get a current impression of the project's status. Appropriate tools which are used in the project for developing Caroline are:

- Specifying story cards: Microsoft Powerpoint.
- Modeling algorithms: UML tools.
- Realization: Eclipse for C++ (CDT), MATLAB/Simulink.
- Version and configuration management: Subversion.
- Unit testing: CxxTest [20].
- Requirements testing: System simulation using scenario-based executable test drive specifications.
- Deployment: Unix-based scripts developed within the project combined with FSVS.
- Software build and integration process: Make/QMake/Cook.
- Milestone planning and tracking: Trac [8].
- Bug tracking: Trac.
- Knowledge management: Wiki provided by Trac.

Our software and systems engineering process relies on a variety of software development tools and some customizations and extensions to combine these tools and to optimize our tool chain. As mentioned before, elements of Extreme Programming, like the test first approach, common code ownership, pair programming, and continuous integration are the basis for an efficient and successful development process. Reducing integration activities to nearly zero through the continuous integration principle is one key element in our development process. As described earlier, this implies that every developer integrates his work frequently and is disciplined in using the controlled source code repository based on the version control system.

Hence, Subversion manages nearly everything necessary to build the project to ensure self-containedness of the entire project. From install scripts, property files and test scripts, up to IDE configurations, the repository contains and provides all project-dependent data. This enables the team to fully build the system from just a checkout on a machine with only a minimum amount of pre-installed software like some third party libraries. Moreover the use of Subversion and the software build tools allow us to setup the fully automated build process, including test runs, which acts as a monitor to the repository needed for quality assurance.

With this approach we are able to find errors quickly and fix them as soon as possible. Triggered by every commit against the repository, central servers start to check out the project sources and initiate an entire build. This is the point where software construction tools come into play. On the developer site these tools help to speed up the build process and at the same time ensure a consistently built result by analyzing changes and dependencies to identify what really needs to be rebuilt. On the server side it allows us to build alternative targets for different forms of use, so we are able to run a system build with or without test code. Furthermore, the testing system separates time consuming high level tests by detaching the complete automated test run to be done in a parallel manner on different servers. Thus, whenever a developer checks in a new version of the software the complete automated set of tests is run.

Feedback of the automated and continuous build and test process is sent to the developer by notification through email and through publication of the build and test results on the project specific portal web site using Trac. This greatly improves responsiveness of the test system. Problems do not remain undetected for a long period, and in only a short time the fault is bounded to a small portion of changed code. Efficiency of the development process as well as a responsible and self-disciplined form of development are effectively assisted by the tooling and testing infrastructure.

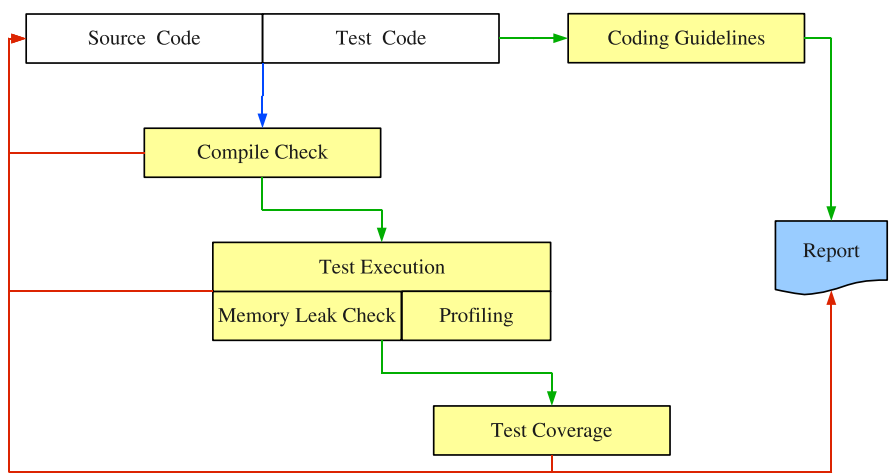

Fig. 4: Multi-level test process: Obviously, the first stage contains the compilation check, followed by a memory leak check which is executed during the test run to find potential memory leaks or NULL pointer access. After executing the test cases, their coverage related to the source code is computed. Here, different coverage criteria like statement coverage or path/branch coverage can be defined and checked. These coverage criteria indicate where the developer should add or modify existing test cases to fulfill the required coverage level. After fixing bugs, memory leaks, or adding more test cases, the algorithm can be optimized using profiling tools. Finally, the code can be checked using a given coding guidelines definition. All results are aggregated into one single report. This multi-level test process can stepwisely executed manually by the developer or completely unattendedly and automatically using the continuous integration system.

Fig. 4 shows the workflow of our multi-level build process. The fully automated process for checking the software quality consists of several consecutive steps. Starting with a compile check, compilation as well as syntactical conflicts are detected pretty early. However, it is expected that code which cannot be compiled is never checked into the version control system.

To automate tests, we are using a light-weight and portable testing framework for C++, called CxxTest. During the test run, the memory checker Valgrind searches for existing and potential memory leaks in the source code. An additional tool from the GNU compiler collection named GCov is used to report the test coverage of the source code. While running the test code it counts and records executed statements. The intent is to implement test cases which completely cover identified critical parts of the software.

Tests are usually implemented directly in C++. Experiences have shown that for a disciplined test definition in a project, it is very helpful that the implementation of tests is done in the same language. This enables an integrated test development process and avoids the hurdle of learning another testing notation. The failure of any single test case causes the complete build to fail and immediate feedback is given to the developers. In order to check real-time properties which are of course necessary for a timed execution in a traffic situation, a step for profiling is done to check the consumed computation time.

For an initial phase, we found it helpful to add another step to the build process, which checks some compliance rules for the coding guidelines. For example, it analyzes the source code for appropriate definitions of names of variables and classes, checks depth of inheritance, number of attributes, sizes of method bodies, appropriateness of indentation, existence of comments, and the like. But moreover, even complicated guidelines like visibility rules through inheritance for example can also be checked. This step can be automated as well [4].

The process outlined above is not only working automatically at every check-in, but can also be executed manually by every developer. When starting a commit cycle the developer first updates his working copy, runs all tests, and then commits his changes only if everything builds well and all tests ran without errors. This leads to a double check, both before the developers commit and automatically at the server side by the master build process. This is reasonable because there is always a chance that the repository was not properly updated by the developers.

As a result of this approach, developers can easily rerun tests and detect many bugs or inconsistent enhancements locally and rather quickly. Thus, fixing discovered bugs is done rapidly and we have a stable and properly working system almost all the time. The benefit is that everyone shares a stable base to start development. Frequent commits, usually more than once a day, guarantee that re-integration of newly developed or enhanced code does not take long and we usually have little integration overhead.

# 5 Application of the Software Engineering Process in the CarOLO Project

As outlined in [2], we are using the system simulation not only for interactive development of an artificial intelligence module in the CarOLO project but also as part of an automated tool chain running on our servers over night. The main goal behind this application is to ensure a certain level of software quality. Therefore, we have extended the test first approach mentioned in Sec. 4 by using the system simulation as shown in Fig. 5. We call this approach the "simulate first approach".

Almost any development process starts with analyzing the requirements documents provided by the customer. In the CarOLO project, we used the DARPA Urban Challenge documents to understand the requirements. These documents contain mostly abstract and non-functional definitions for the autonomously driving vehicle. On the one hand, these requirements were rather stable–even though they were occasionally changed. On the other hand, they were rather vague and left open a lot of options for possible traffic situations, weather conditions, forms of roads, etc. Three big phases lead the project from an initial setting through the first application (end of phase 1), through the site visit (end of phase 2) to the National Qualification Event (NQE) and Final Event (phase 3) as already described in Sec. 3. Actually, in any good development project, there is a final phase with a retrospective and a discussion of the insights, the gathered knowledge and experience, and a consolidation of the developed software. This also includes the identification of reusable parts of the software. Each of the first three phases is broken up into several small iterations.

In every iteration a new task dealing with a coherent requirements group is chosen by the development team, prioritized in the product backlog, and defined using the Scrum process for agile software engineering as mentioned in Sec. 2. These requirements are refined into the already discussed story cards and scenarios are designed for both a virtual test drive and a real vehicle test for evaluations using the completely equipped car. This early definition of the virtual test drive forces developers to clarify general parameters and conditions before starting their implementation. The result is an *executable test drive specification* that tests all requirements to be implemented. Now, the implementation of the system and the virtual tests can run in a parallel manner.

In the testing phase, after designing a virtual test drive, the availability of necessary validators is checked. If there is a condition which is not handled yet by a required metric, an appropriate validator is implemented. As mentioned earlier, these validators are the base for automated test runs, which are necessary for the continuous integration of the complete software system and therefore a vital part of a consistent software system engineering process.

The newly implemented test cases are grouped together in a test suite and form an *executable specification* of the virtual test drive. The new test suite is finally integrated in the tool chain. None of the old test suites should fail and only the new one should not pass. With these failed tests, the implementation of the new artificial software functions begins. In small iterative development steps the software module is extended for fulfilling every test case of the new test suite.

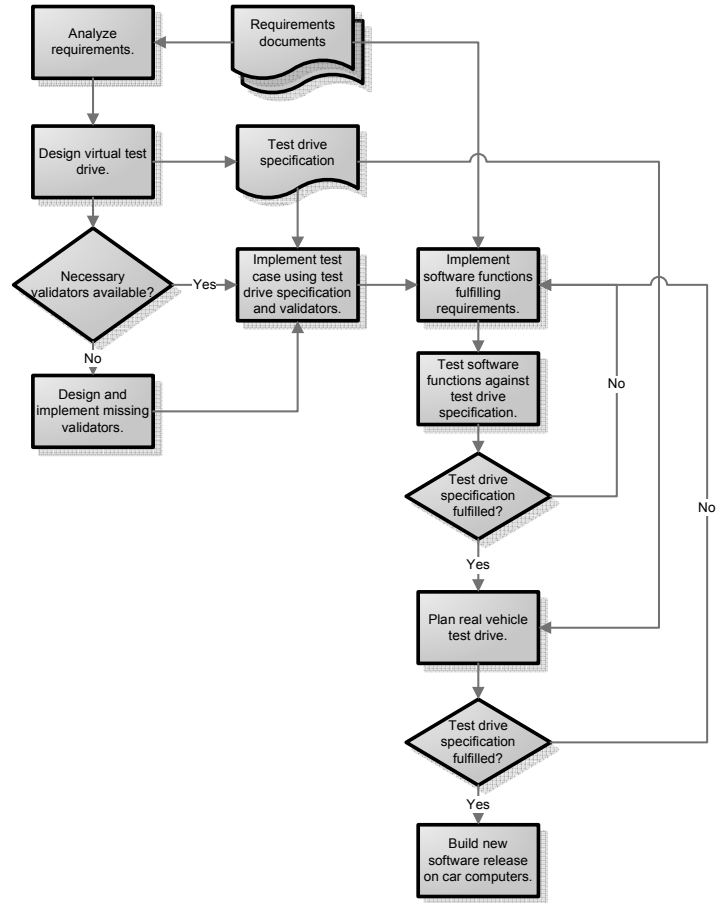

Fig. 5: Simulate first approach inspired by the unit testing approach: Here, the customer's requirements are part of the technical software development process because they are the basis for defining scenarios to be used in interactive simulations. For setting up unattended simulations, not only the customer's needs are necessary, but even more his acceptance criteria have to be discussed to define so called validators which continuously supervise a running system in a system simulation to evaluate whether it fulfills the customer's needs or not.

Although the above sounds like a normal test first approach, there are a number of differences. First of all, the test suite which captures the high level requirements for handling traffic usually does not change the signature of the overall system. The reasoner as well as the system simulation have stable interfaces and need only changes in its behavioral algorithms. So we do not need to define interfaces before tests can be defined. And second, these high level tests are black-box and do not rely on the internal structure of the code. However, for a thorough test of the reasoner, it is also helpful to add more system simulation based tests after the code is implemented to check the states and transitions between them that occur in the reasoner as well as traffic situations that the reasoner has to handle.

As with usual test first approaches, these small steps are iterated until the complete test suite is satisfied. After completing the implementation, the new intelligent software function will fulfill the requirements in the virtual test drive. Subsequently, the test drive is planned and executed using the real vehicle if applicable.

If the test drive is successful the new software can be released and marked as stable. After the test drive, usually optimizations must be implemented, bugs must be fixed, and technical issues or misunderstandings from the requirements documents must be fixed. Before the code is modified, the system simulation is used again to extend the test suite in such a way that the error becomes visible under a virtual test and then the software is modified. These virtual test drives can be repeated at nearly no cost and help the team to develop quickly and on time the necessary software functions.

In a second variant, we have enhanced our system simulation environment in such a way, that multiple car instances can drive in the same world model. Using multiple instances allows running several intelligent cars together with each other. On the one hand this is a good way to investigate artificial learning of optimal driving behavior in a potentially dangerous world. On the other hand, we can handle racing conditions by running multiple instances of the artificial intelligence that start from the same position in the world data with the same mission, i.e. have to drive the same route. Starting multiple instances with the same mission data allows us to understand and compare the performance of several versions of our intelligent car. It is also possible to measure the stability of the intelligent algorithms over time when using the same software revision in slightly enhanced courses. Furthermore, it is possible to watch the virtual Caroline becoming an even more optimal driver based on the increasingly optimized revisions of the artificial intelligence modules.

## 6 The Hesperia Framework – An Integrated Approach for a Virtualized Software Engineering

In the CarOLO project, we developed and used a specialized software framework for system simulations as outlined in [2] and [13]. However, this framework had some design drawbacks. One important issue was the chosen communication concept. In that software framework, we have chosen TCP for reliability purposes to send data from one component to another. TCP allows a stream-oriented, directed, and most important a synchronized communication. These characteristics are suitable for different contexts; however, in the context of autonomous driving, where several components are sending compact messages with a high frequency, another protocol is more suitable.

Moreover, TCP does not allow to broadcast messages only once for several receivers. Hence, the same message is sent twice or even more often depending on the number of receivers and thus, the data is redundant on the network which might causes packet collisions, consumes valuable bandwidth, and even worse costs transmission time.

Furthermore, the communication based on TCP is directed. Thus, all modules are knowing their communication counterparts. Therefore, the system architecture must be defined a-priori in a very mature manner to avoid frequent changes to communication paths between all modules. And even more, if another component C needs the same data from a component A which is already sent to a component B, component A must be modified to fulfill component C's needs.

Moreover, the software framework used in the CarOLO project did not provide a formal specification to define scenarios with different situations in a machine processable manner. Thus, the setup of virtual test scenarios was not supported well.

Therefore, these conclusions and results yielded the development of an integrated approach for a virtualized software engineering [3]. This approach is called "Hesperia"[1] realizing a multilayer software framework which is outlined briefly in the following.

### *6.1 System Architecture of an Autonomous System*

In Fig. 6, the general system architecture of an autonomous is shown. Such a system which is embedded in the system's context consists of three main layers: Perception layer, decision layer, and action layer.

The gray inner box depicts the complex embedded system itself consisting of the aforementioned three processing layers. The system itself can be supervised using tools from the support layer. Some important tools are *recorder* for recording *non-reactively* data exchanged by the running components and *player* for replaying the data. Moreover, *monitor* is available to visualize complex data flows like the perceived sensor raw data or a computed trajectory to be driven by the vehicle. This component also does not interfere at all with the three aforementioned layers compared to the software framework developed in the CarOLO project.

For allowing interactive as well as unattended system simulations, a layer for *closing the data processing loop* is necessary. This layer is called *virtualization layer* due to its main task to simulate either missing components of the real system like sensors or the modify a system's context like moving other vehicles using a mathematical model.

---

[1] The name "Hesperia" is derived from a town in California where the team CarOLO was accommodated.

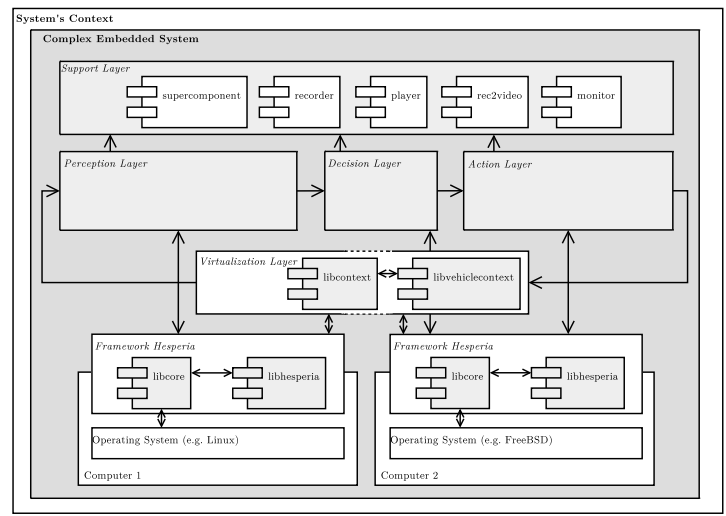

Fig. 6: System architecture for a complex embedded system with sensors and actuators. In the center of this figure, the main processing components are depicted: Perception, Decision, and Action. The first layer perceives the system's surroundings for extracting relevant features or abstracting from the incoming raw data. The second layer analyzes and interprets the preprocessed data for deriving a set of actions which are processed by the last layer by using control algorithms for example. For inspecting the system non-invasively the support layer offers possibilities to monitor a currently running system or to record and replay previously captured data. For setting up interactive or unattended system simulations, the previously described processing chain must be closed. This is realized by the virtualization layer provided by the framework Hesperia.

Technically, any component can be virtualized without modification using the library *libcontext* which is responsible for controlling the overall system time and the entire communication. Thus, the running system under test is decoupled from the real system time and can be run faster or slower if necessary. Thus, this library as part of the software framework Hesperia is the enabling technology for re-using unmodified components in unattended and repeatable system simulations.

For providing suitable mathematical models of the driving behavior of vehicles, the library *libvehiclecontext* was developed. This library can be used for interactive as well as unattended system simulations.

### 6.2 Modeling the System's Context

For defining repeatable and meaningful system simulations, a formal description of scenarios and situations is necessary. This input data can be derived for example from the aforementioned story cards. In Fig. 7, an excerpt of the formal description of a scenario is depicted.

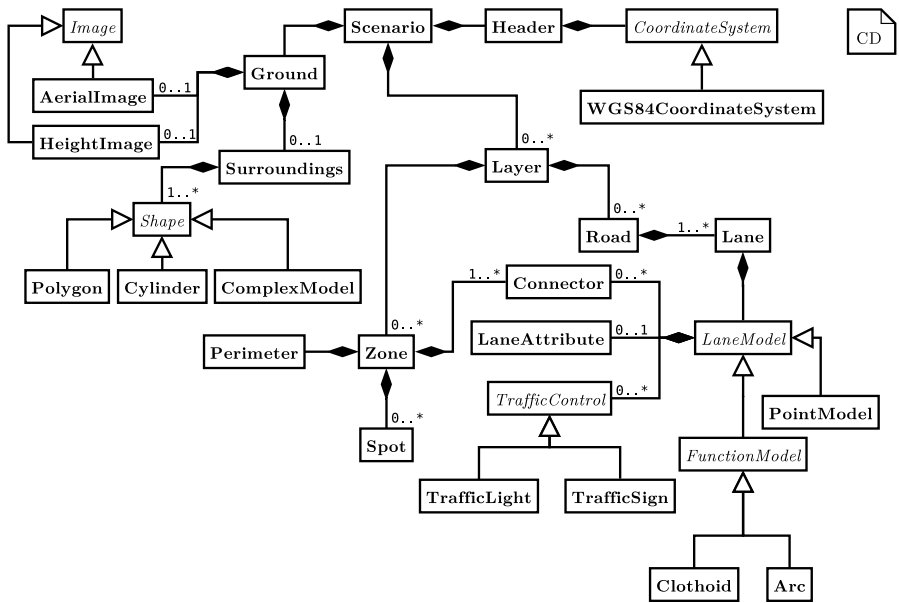

Fig. 7: Excerpt from the formal description of a scenario. This UML diagram is the base for a domain specific language.

This formal description consists of more than 25 entities for describing the stationary system context and nearly 20 entities to define dynamic elements like other vehicles with their associated behavior. For using this formal description in a machine processable manner, a domain specific language (DSL) was derived. The DSL was designed using experiences from the 2007 DARPA Urban Challenge as well as conditions from a context of a vehicle which is driving automatically on highways [23].

For optimizing the DSL itself, its first implementation was made using the MontiCore framework for developing and using DSLs [10]. Thus, the language could be simply enhanced and modified to be suitable for use in automatable system simulations. Furthermore, a graphical editor as shown in Fig. 8 based on the Eclipse Rich Client Framework [19] was realized easily re-using the Java sources generated by the MontiCore framework [17]. In that figure, a traffic situation at an intersection is shown.

For processing the DSL in the software framework Hesperia, a C++ lexer and parser are necessary. To avoid additional tooling in the software build process, which may be error-prone or cause inconsistencies between the generated classes from the DSL and the classes using these generated ones, a compile-time generated implementation was chosen based on Boost Spirit [5]. Here, the definition of the DSL is provided by a C++ template specification which is read by the C++ compiler itself to generate the necessary classes for the non-terminals and terminals of the language.

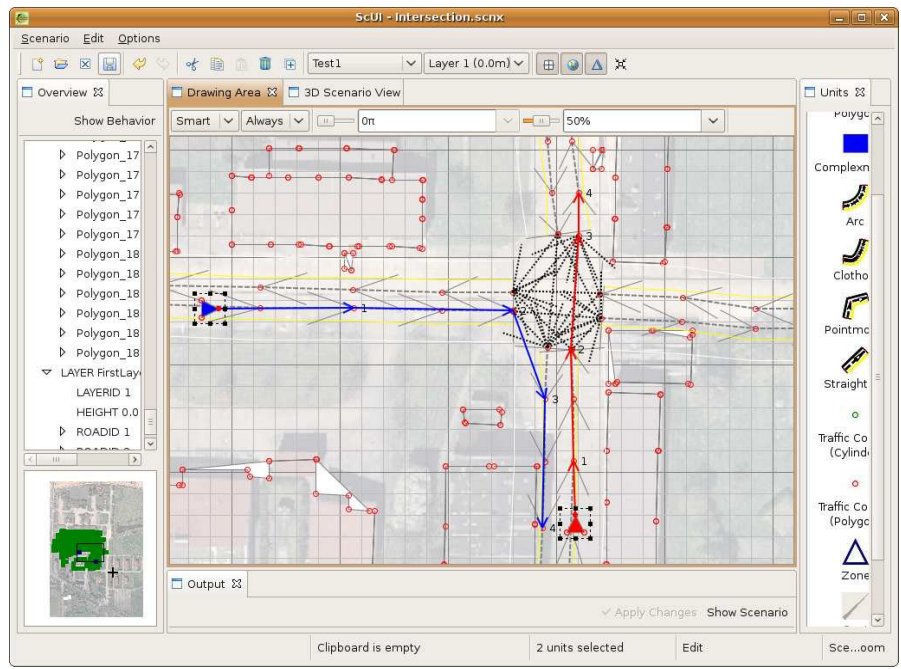

Fig. 8: Graphical editor for using the DSL realized with the Eclipse Rich Client Framework. In the lower left corner, a small map of the scenario from the bird's eye view is shown. On the left hand side, a hierarchical representation of the DSL's instance for the currently edited scenario is shown which can be modified by the user. In the center of this figure, the main editor view is depicted which shows a situation at an intersection modeled for two vehicles. On the right hand side, a set of elements is available which can be added to the scenario by simply using drag and drop.

This Boost Spirit framework itself is wrapped in the software framework Hesperia to encapsulate the access to the classes of the abstract syntax graph (ASG) as well as to handle errors in the given DSL's instance appropriately.

The DSL is a core part of the software framework Hesperia for defining interactive or unattended system simulations. Therefore, not only abstract elements like lane markings or traffic rules can be defined; even more, complex and detailed 3D models from popular 3D modeling applications can be simply added to a scenario. Thus, not only a realistic appearance for the user can be realized but these models are also used to generate realistic sensor raw data for example as described in the following.

*6.3 The Framework Hesperia*

In Fig. 9, all core components of the software framework Hesperia are shown. First, one of Hesperia's core concepts is to decouple a module's dependencies to third party supplied libraries. Therefore, inspired by the *design-by-contract* concept, only interfaces of a module's required existing algorithms are available for the module. The library supplied by a third party is entirely wrapped in lower layers allowing a transparent exchange if necessary. For example, for fast data storage, the BerkeleyDB is transparently available for all modules; for processing images as well as matrices, the OpenCV library is wrapped and a generic interface is exported to higher layers.

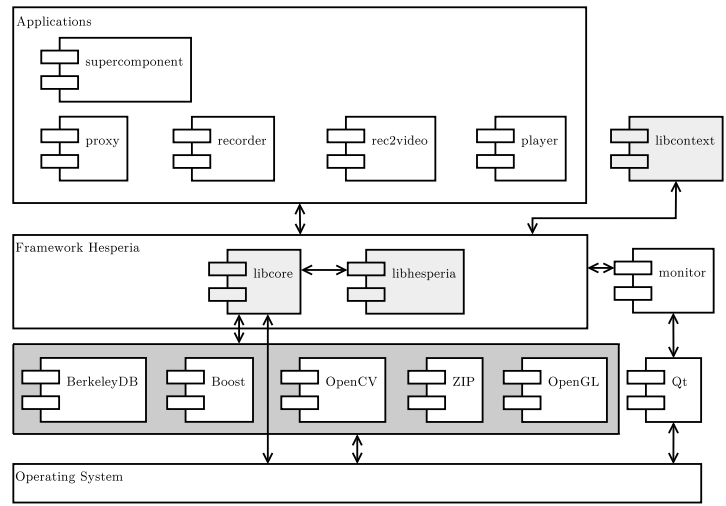

Fig. 9: Components of the software framework Hesperia.

In *libcore*, rudimentary system services are implemented and intensely tested using CxxTest. These services include concurrency for realtime and non-realtime services, I/O handling for URLs, and the basic communication concept *ClientConference*. A *ClientConference* is an enhanced wrapper around the UDP multicast protocol. Using UDP multicast as the core communication concept, all data is sent only once regardless how many recipients are listening.

Furthermore, the communication is realized using untyped "conferences" but typed messages called *Containers* which contain one serialized object. Using this concept, the receiver can use different thread-safe data storages like FIFOs, LIFOs, or key/value maps provided by *libcore* to receive data using the listener/observer pattern. Furthermore, these data storages can be simply reused to filter and combine different data like position data and features extracted from images.

The serialization of data structures is realized in a query-able manner. Therefore, a class' attributes-dependent identifier is computed at compile time. This identifier is used to access a class' attributes in a random manner during deserialization. Thus, changing the order or adding new attributes to a class do no break existing components; furthermore, these existing components do not need to be re-compiled.

For supporting the development, some additional components are provided with Hesperia. The most important component is *supercomponent* which is mandatory for every *ClientConference*. This component tracks the lifecycle of each running component and delivers the module-dependent configuration data using the *Dynamic Module Configuration Protocol* (DMCP) which is inspired by the well-known DHCP. Thus, the configuration data can be modified centrally on the supercomponent's node and is consistent for a complex embedded system which is distributed automatically to several computing nodes.

As mentioned before, components for recording and replaying data are also available. These component can transparently record all transferred data of a running system without interfering with other components by simply joining a running *ClientConference*. For producing videos to illustrate a traffic situation, the component *rec2video* can be used. This component reads a previously recorded session and generates frames to render a video file.

### *6.4 Generating Synthetic Sensor Raw Data*

As mentioned above, the software framework Hesperia provides libraries to generate synthetic sensor raw data. On the example of a single layer laser scanner, results are shown in Fig. 10.

For producing this raw data, the formally specified system context is used provided by the software framework Hesperia. Therefore, algorithms also described in [3] were integrated in the software framework which is outlined briefly in the following. The algorithm bases on a scene which can be rendered using OpenGL. Therefore, the system context is transformed automatically into a renderable scene representation using the visitor concept which traverses the ASG produced by the DSL's parser.

After generating the scene graph, a camera which is looking into this scene is defined for each single layer laser scanner. This camera is positioned to the corresponding virtual mounting positions of the scanners.

Following, a so-called projector is created for each camera. This projector is used to project a texture into the scene which describes the shape of the scanning line. Using a specialized shader program which is executed on the GPU, the distances to the camera, where the line hits the first visible object, are encoded into the resulting rendered image.

The final step of this algorithm is to analyze this synthesized image by interpreting the encoded distances to transform them into the local scanner coordinate

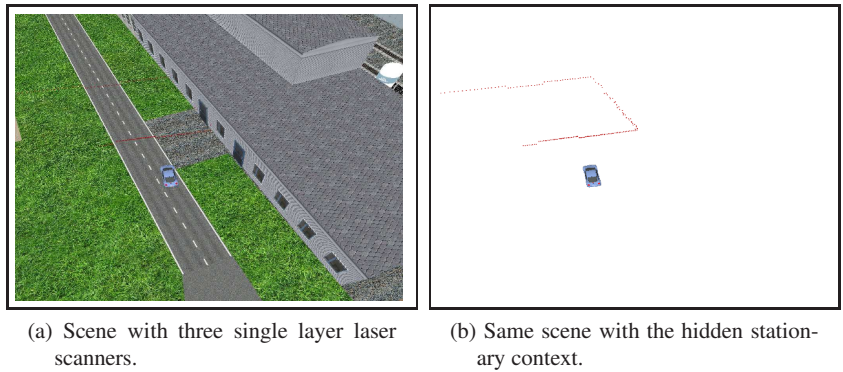

(a) Scene with three single layer laser scanners.

(b) Same scene with the hidden stationary context.

Fig. 10: The software framework Hesperia can be used to generate synthetic sensor raw data for a single layer laser scanner and camera systems. In this example, three independent scanners are defined: Two pointing to 8m and 15m in front of the vehicle which are scanning orthogonally and one scanner which is scanning in a parallel manner to the driving direction.

system. Then, the resulting coordinates can be serialized and distributed to imitate a sensor's protocol.

A similar principle without the shader programs can be used to simulate a simple color or gray-level camera. Therefore, the current rendered image "captured" by the OpenGL camera is read back from the GPU and provided as input data for vision algorithms.

### 6.5 Automating System Simulations

As mentioned before, *libcontext* and *libvehiclecontext* can be used to automate the execution of system simulations. As outlined in Sec. 5, the customer's needs must be available to define acceptance criteria for the software under test. Compared to unit tests, acceptance criteria define the set of conditions when to evaluate a test scenario to passed or to failed.

The software framework Hesperia was successfully applied to a research and development project for the development of an autonomously driving vehicle which should navigate on a given digital map. The project was carried out at the University of California, Berkeley together with the RWTH Aachen University from June to August 2009. Therefore, three validators for continuously evaluating the software in a purely virtualized system's context were developed.

The first validator called *DestinationReachedReporter* continuously supervises the current position and orientation of the vehicle. It returns true, if the vehicle finally reached its overall destination. Therefore, this validator is registered at the virtualization layer at *libcontext* as a read-only listener for data sent in the *Client-*

*Conference*. Thus, the entire system does not need to be modified and can instead be inspected non-invasively.

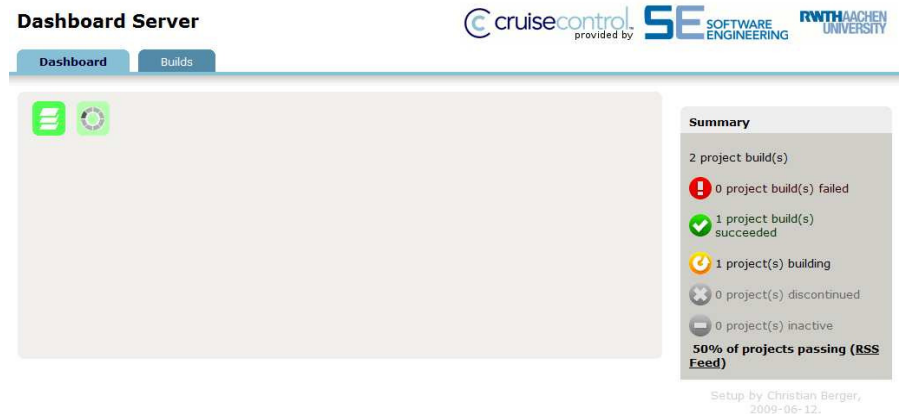

Fig. 11: Quality assurance with the continuous integration system CruiseControl. Using continuous integration, the same source code and test cases can be checked for different compiler options as shown in the figure, or even more for different platforms like Linux and FreeBSD. Thus, the developers can concentrate themselves on the core development of the algorithms and must only modify or correct the code if one integration server reports an error for a specific platform.

The second validator is called *ShortestRouteChosenReporter* and evaluates to true, if the vehicle has chosen and driven the shortest route between two given waypoints. Therefore, it computes the shortest route using the digital map provided by the DSL which describes the stationary context. The result of this computation consists of a list of consecutive waypoints to be passed. During execution, this validator continuously supervises the vehicle's position and orientation and returns true when finally all waypoints were successfully passed.

The third validator, which is called *DistanceToRouteReporter*, continuously supervises the position and orientation of the vehicle and measures its distance to a given route. If the distance was finally never greater than a given threshold, this validator returns true.

All the aforementioned validators were combined in different test cases to define criteria for quality assurance using the unmodified CxxTest unit test framework; other validators are described in [3]. Thus, these test cases could be easily integrated with CruiseControl [7] as shown in Fig. 11.

Compared to the CarOLO project where a set of self-written scripts were used to realize a continuous integration system, CruiseControl could be applied without modification. Moreover, older test runs can be still accessed through CruiseControl's web front for further inspection. For example, for two test cases to evaluate the aforementioned navigation algorithm, already 12MB of results in XML are gener-ated per execution. Thus, appropriate post-processing such as XSLT must be applied to aggregate the information and to create a valuable report for the developers.

# 7 Related Work

The approach described in the first part of this article has been used to develop the autonomous driving vehicle "Caroline" for the 2007 DARPA Urban Challenge. In the CarOLO project a 2006 VW Passat station wagon has been modified using car computers and many sensors for perceiving and understanding the environment and actuators for controlling the car by software. In [2], the simulation approach developed for the competition is presented. However, and already discussed in Sec. 6, the solution developed for "Caroline" had some drawbacks. Thus, as presented in the second part, the software framework Hesperia written entirely from scratch allows a more convenient and repeatable way to design and execute system simulations [3]. Besides the results and experiences from the 2007 DARPA Urban Challenge, experiences from a research project to develop an automatically driving vehicle for highways found the base for the development of the software framework Hesperia.

Compared to the approach presented in this paper, similar approaches for simulation purposes are the CarMaker Vehicle Simulation by IPG Automotive GmbH and VEDYNA, [11] and [18], numerical simulations of full car dynamics with interfaces to MATLAB/Simulink. Both try to ease the development and integration of vehicle controllers. However, they do not support the consistent and integrated simulation of software architectures and software components using an embedded domain specific language including the re-use of available 3D models.

A similar approach as provided by *libcore* from the software framework Hesperia is realized in the Automotive Data and Time Triggered Framework (ADTF) as outlined in [14]. The ADTF can be used to model a directed graph reflecting the data flow through a set of processing modules. The communication is realized using so-called channels, which themselves are typed but which can carry arbitrary typed data in principle contrary to the approach realized in the software framework Hesperia which relies solely on typed messages instead. Furthermore, no support for a formally specified, consistent data model is provided.

Additionally to the aforementioned ADTF, the toolkit Virtual Test Drive is developed to manage previously recorded raw sensor data or to synthetically generate required input data to perform SiL-, HiL-, ViL-, or DiL-simulations [12]. Compared to the software framework Hesperia, an approach to generate sensor raw data for a single layer laser scanner for example is still not available yet.

A similar approach to the software framework Hesperia and the tool-suite Virtual Test Drive is provided by TNO PreScan [9]. This software can be used to support the development of so-called pre-collision driver assistance systems. Contrary to the software framework Hesperia, the synthetic generation of sensor raw data for a single layer laser scanner for example based on popular 3D models is not supported and no formal and integrated DSL is provided for the developer.

Another approach is provided by a tool from IAV [15]. This tool generates synthetic raw data for arbitrary sensors. Therefore, the user models in a 2D manner the characteristics of a specific active sensor like a field of view (FOV), a maximum distance, and some error noise. Then, the software computes preprocessed sensor data which would be provided by the ECUs of a specific sensors. Contrary to the software framework Hesperia, only open-loop data generation is possible. Thus, no resulting sensor data in interaction with other vehicles can be generated. Furthermore, the sensors' models are only so-called visibility models which do not imitate the actual measurement principles.

## 8 Conclusion

Intelligent driving assistance functions need a detailed understanding of the vehicle's surroundings, of the driving situation, and the traffic rules and regulations as well as a sufficient knowledge about the physics of cars to fulfill their tasks. In the end, an intelligent driver assistant must be able to drive on its own. Thus, the 2007 DARPA Urban Challenge was a great opportunity to foster this area of autonomous vehicles and intelligent driving assistants. Developing this kind of complex software needs an innovative, agile development process that is compatible with the overall system consisting of a standard car, such as a VW Passat, sensors, actuators, and a number of computers suitable for automotive use.

For an efficient and stringent development project, a number of actions has to be taken, including an iterative development in small increments, early bug-detection and bug-fixing, stable version and configuration management, a solid architecture which embraces automated tests at any level of the software architecture, and most of all, a thoroughly designed test infrastructure. Evaluating a software's quality includes tests of the entire system, but for efficiency reasons it is important to test as much as possible while focusing on the subsystem under test. Thus, individual methods and classes are tested in the same manner as the whole reasoner. The test architecture allows us to fully extract the reasoner into virtual, simulated traffic situations, and allows to check the car behavior in various traffic situations efficiently. Automation of these tests allows us to (re-)run tests as desired at least every night or for every commit to the versioning system.

There are many complex traffic situations, let alone junction layouts and various possibilities of behavior of other cars, that it is inevitable to run many tests in a simulated environment. The system simulation is rather general and will be usable for testing and interactive simulation in other contexts as well, e.g. it can be combined with HiL-tests.

The approach used and the results gained in the CarOLO project show that autonomous driving is still a few years ahead, but also that efficient development of complex software in combination with the overall system is possible if the development process is disciplined, yet responsible, agile, and assisted by appropriate modern tool infrastructures.